\documentstyle[prl,aps,amssymb,graphicx,twocolumn]{revtex}

\begin{document}
\title{Hanle effect in Coherent Backscattering}
\author{G. Labeyrie$^{\ast +}$, C. Miniatura$^{\ast }$, C. A.\ M\"{u}ller$^{\ast }$%
, O. Sigwarth$^{\dagger }$, D. Delande$^{\dagger }$ and R. Kaiser$^{\ast }$}
\address{$^{\ast }$  Laboratoire Ondes et D\'{e}sordre, 
FRE 2302 CNRS, 1361 route des Lucioles, F-06560 Valbonne.\\
$\dagger $ Laboratoire Kastler Brossel, Universit\'{e} Pierre et Marie Curie, 
4 Place Jussieu, F-75005 Paris.\\
$^{+}$ e-mail : labeyrie@inln.cnrs.fr\\}
\date{\today}
\maketitle

\begin{abstract}
We study the shape of the coherent backscattering (CBS) cone obtained when
resonant light illuminates a thick cloud of laser-cooled rubidium atoms in
presence of a homogenous magnetic field. We observe new magnetic
field-dependent anisotropies in the CBS signal. We show that the observed
behavior is due to the modification of the atomic radiation pattern by the
magnetic field (Hanle effect in the excited state).
\end{abstract}

When a multiply scattering medium is illuminated by a laser beam, the
scattered intensity results from the interference between the amplitudes
 associated with the various
scattering paths; for a disordered medium, the interference terms are washed
out when averaged over many sample configurations, {\it except} in a narrow
angular range around exact backscattering where the average intensity is
enhanced. This phenomenon, known as coherent backscattering (CBS),
is due to a two-wave constructive interference (at exact back-scattering)
between waves following a given scattering path and the associated reverse path, where
exactly the same scatterers are visited in the reversed order \cite{CBS}. The maximum
enhancement is obtained when the amplitudes of the interfering paths are
exactly balanced. For a convenient choice of polarization, time-reversal
symmetry directly implies the equality of the interfering amplitudes. More
generally, the interference phenomenon is very robust and qualitatively
insensitive to most characteristics of the sample and illuminating wave.

However, applying a magnetic field on the sample breaks the time-reversal
invariance. It was predicted \cite{Golub}, then experimentally observed 
\cite{CbsBexp} and theoretically studied \cite{CbsBth,Lacoste} that it results 
in a decrease of the CBS enhancement as well as some rather complicated
behavior of the cone shape. In our current understanding of CBS, two
ingredients are essential : the individual scattering event, characterized
for example by the radiation pattern of each scatterer (this is the single
scattering ingredient), and the propagation in the medium between scattering
events (this is the ''average effective medium'' ingredient). Of course, these
two ingredients are not independent since the optical theorem links the propagation 
in the medium to the individual scattering. In the
presence of a magnetic field, both ingredients can be affected. For the
propagation, this is well known under the name of Faraday effect (and 
the Voigt or Cotton-Mouton effect) and can be described by the
modification of the complex refractive index by the magnetic field. The
magnetic field-induced variation of the radiation pattern is much less
studied, because it is very small in usual magneto-optically active
materials; it is responsible for the ''photonic Hall effect''
predicted by van Tiggelen \cite{PHEth} and later observed experimentally 
\cite{PHEexp}. In this paper, we show a novel situation where it is
experimentally and theoretically possible to discriminate between the 
two ingredients and where the modification of the radiation pattern
dominates the propagation effects.

In our experiment, we analyze coherent backscattering of resonant light by a
dilute gas of laser-cooled rubidium atoms. Some features of this medium
differ markedly from those used in previous CBS experiments. First, the cold
atomic cloud constitutes a monodisperse sample of point scatterers, highly
resonant in the vicinity of the optical transition. This implies a dramatic
increase of the scattering cross section at resonance, but also a great
sensitivity to any external perturbation like a magnetic field. Typically, few
Gauss are enough to bring the atom completely off resonance, in sharp
contrast with previous studies where Teslas were needed to induce
significant effects \cite{CbsBexp}. Concurrently, a giant Faraday effect is
observed in the cold atomic cloud \cite{Faraday}. Another important
difference with classical systems is the presence of a quantum internal
structure, which has been shown to strongly reduce the coherent
backscattering interference \cite{CBSat}. Obviously, the addition of a
magnetic field will affect the contrast of the CBS cone in various ways, and
we have indeed observed new behaviors that will be described somewhere else. In the present
Letter, we focus on the shape of the cone rather than its intensity.

The experiment has been described elsewhere \cite{CBSat}. We use the $%
J=3\rightarrow J^{\prime }=4$ transition of the D2 line of Rb$^{85}$
(wavelength $\lambda =780$ nm, natural width $\Gamma /2\pi =5.9$ MHz) 
where $J$ here denotes the hyperfine angular momentum. We
show in Fig. 1 some CBS cones obtained in the linear polarization channels,
for different values of the magnetic field and laser detuning. After scaling and
substraction of the incoherent background (average intensity in the wings of
the peak), we plot the detected far-field intensity isolines. 
The horizontal and vertical
coordinates correspond to the two azimuthal scattering angles.
 The magnetic field
${\mathbf B}$ is orthogonal to
the plane of the figure, like the
backscattered light wavevector ${\bf k}_{\text{{\bf b}}}$ which
points towards the reader. The incident light
polarization is parallel to the vertical axis of the figure. Since we will
care about the sign of magneto-optical rotations, we have to precise our
conventions. A positive magnetic field is parallel to the incident light
wavevector ${\bf k}_{\text{{\bf i}}}(=-{\bf k}_{\text{{\bf b}}})$ with the
same orientation. A positive angle corresponds to a counterclockwise
rotation when ${\bf k}_{\text{{\bf i}}}$ points towards the observer.

\begin{figure}
\begin{center}
\includegraphics[scale=1.0]{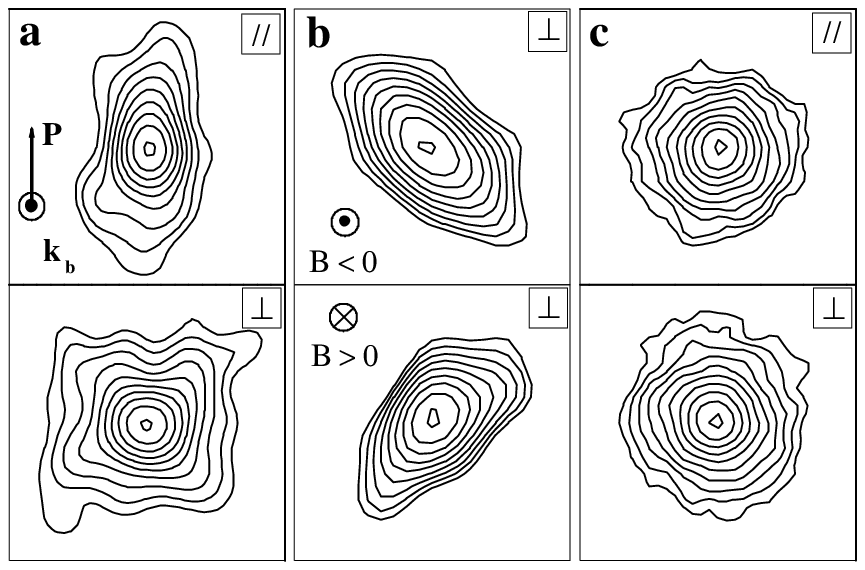}
\end{center}
\caption{Experimental CBS cones. The far-field backscattered intensity is 
plotted
as a function of the two azimuthal scattering angles (maximum intensity at 
center; we use a logarithmic scale for the isolines for a better
visualization of the wings where the anisotropy is more pronounced;  
the lowest isoline corresponds to
roughly 10\% of the peak value). The total angular range is 2 mrad. ${\bf P}$ denotes
the incident polarization.  
(a) $B=0$, top: $lin\,{\parallel }\,lin$, bottom: $lin\,{\perp }\,lin$; 
(b) both cones in $lin\,{\perp }\,lin$, top: $ B=-8$G, bottom: $ B=+8$G; note
the $90^{\circ}$ flip when $B$ changes sign. 
(c) $B=+10$G and $\delta _{\text{L}}=-2.6 \Gamma$,
top: $lin\,{\parallel }\,lin$, bottom: $lin\,{\perp }\,lin$.}
\end{figure}

For resonant light (laser detuning 
$\delta _{\text{L}}=\omega _{\text{L}}-\omega _{\text{A}}=0$ where
$\omega _{\text{L}}$ and 
$\omega _{\text{A}}$ are the laser and atomic frequencies respectively) and
zero magnetic field $B=0$, we obtain in the $lin\,{\perp }\,lin $ 
channel a cushion-shaped cone with a four-fold symmetry shown in
Fig. 1(a) (bottom). In the parallel channel (top), the cone has an
elliptical shape with its large axis parallel to the incident polarization.
In the presence of a magnetic field, we observe in the $lin\,{\perp }\,lin$
channel (b) that the cone has now an elliptical shape with an inclination of 
$45^{\circ }$ from the incident polarization, which does not depend on the magnetic
field strength. It is remarkable that, when
the sign of the magnetic field is reversed, the cone flips by $90^{\circ }.$ 
In the $lin\,{\parallel }\,lin$ channel, the shape
of the cone does not significantly change with the magnetic field. When the laser is detuned with respect to
the zero-field atomic transition, the cones in both the 
$lin\,{\parallel }\,lin$ and $lin\,{\perp }\,lin$ channels 
become isotropic, as shown in Fig. 1(c). 

To understand the features observed in Fig. 1, we need to remember that the 
CBS cone shape is closely linked to the distribution of distances between
first and last visited scatterers in the sample. In fact, it is the
Fourier transform of the transverse intensity profile of the scattered
light when the sample is illuminated by a narrow light beam \cite{CbsBexp}. 
Thus, any
scattering asymmetry inside the medium immediately results
in a cone shape asymmetry. For the discussion of the observed effect, 
it turns out (see below) that the
internal structure (Zeeman sublevels) of the atom is not crucial. We thus
first consider the simplest case where the atomic transition is $%
J=0\rightarrow J^{\prime }=1$. The Zeeman effect induces a splitting of the
atomic resonance line in 3 components separated by $\mu B$ where $B$ is the
magnetic field and $\mu /2\pi =1.4$ MHz/G is the Zeeman shift rate. As soon
as the Zeeman shift is comparable to the resonance width (magnetic
field of the order of 4 Gauss), the scattering cross-section of the atom is
strongly modified. For simplicity, we restrict the discussion to a  
resonant excitation $\delta _{\text{L}}=0$. In this case, the response of
the atom to a linearly polarized field is that of a dipole rotated from the 
incident polarization by an angle $\varphi
=\arctan(2\mu B/\Gamma )$, in the plane perpendicular to $B$. 
This rotation is responsible for the
well-known fluorescence dip curve observed in typical Hanle experiments in
the excited state \cite{Hanle}. The effect of the magnetic field on the
average effective medium is also well known: in the direction of
the magnetic field, it induces the Faraday effect, i.e. a rotation of the 
electric field around the magnetic field. For a weak magnetic
field $\mu B\ll \Gamma $, the Faraday angle $\theta _{\text{F}}$ per
mean-free path is : $\theta _{\text{F}}\simeq -\mu B$/$\Gamma $ \cite
{Faraday}. It is crucial to note that it has a sign {\it opposite} to that
of the atomic dipole rotation. As CBS involves both
propagation and individual scattering, the experimental observation of the 
sign of the rotation makes
it possible to determine the dominant effect. In our case, we will show that the
rotation of the atomic dipole leads.

We propose a simple model which explains the observed cone shape behavior, 
taking into account only the rotation of the
atomic dipole. We will restrict the discussion to the case of double
scattering, as it is known to be dominant in our experimental conditions and
because anisotropy effects are expected to decrease for higher order
scattering. Due to the exponential attenuation of light inside the sample,
the first and last scatterers of most paths lie in a ''skin layer'' of
thickness approximately one scattering mean-free path under the surface of
the sample. Thus, we will assume that all propagation takes place in the
transverse plane, orthogonal to the incident wave vector. In addition, we will neglect all
propagation effects (i.e. Faraday and Cotton-Mouton).

Let us now consider an initial polarization vector ${\bf P}$. The first
scatterer radiates a wave at an angle $\Psi $ from ${\bf P}$ towards the
second atom, which then radiates in the backscattering direction where the
field is analyzed along the polarization ${\bf A}$ at an angle $\alpha $
with ${\bf P}$. The angular dependence of the interference term between reverse 
paths is readily given by $\sin ^{2}(\varphi -\Psi )\ \sin ^{2}(\varphi +\Psi -\alpha )$.
This formula (valid for resonant light) allows us to understand most of the
observed behavior of the CBS interference pattern. 
Using a general symmetry argument, one can show that the CBS cone should remain symmetric 
with respect to the bisectors of the incident and detected polarizations directions. 
Indeed, the expression given above is invariant by the transformation $%
\Psi \rightarrow \alpha -\Psi .$ In a polar plot, it exhibits two orthogonal
pairs of lobes along the two bisectors. Each pair has two symmetric lobes,
and the intensity ratio $R$ between the two pairs is $R=\tan ^{4}(\alpha
/2-\varphi )$. Thus, in most cases one lobe pair dominates over the other, yielding a 
cone with an elliptical shape along one of the bisector. When varying the value of 
$\alpha /2-\varphi$ (either by scanning the magnetic
field or rotating the analyzer), the intensity of the dominant lobe
decreases while the smallest lobe grows, until $\alpha /2-\varphi =\pi /4$,
where they are of equal intensity (a four-fold symmetry of the cone is there
recovered). Above this threshold value, the roles of the two pairs of lobes
are interchanged: one thus observes a flip of the CBS cone orientation by 
$90^{\circ}.$ Moreover, in the $lin\,{\perp }\,lin$ channel ($\alpha =\pi /2$), 
$B \rightarrow -B$ flips the cone as observed in Fig. 1(b).  
In our model, the value $\alpha _{f}$ corresponding to the
flip threshold is simply related to the magneto-optical
rotation through $\alpha _{f}=2\varphi+\pi /2$. So far, we have neglected 
all propagation effects. However, note that the Faraday rotation due to 
the propagation of the light through the "skin layer" 
separating the sample surface from the first scattering event \cite{Lacoste} would 
have the same influence on the cone shape as the dipole rotation $\varphi$. As it will be 
discussed later, our results indicate that the dipole rotation dominates in this experiment.

\begin{figure}
\begin{center}
\includegraphics[scale=1.0]{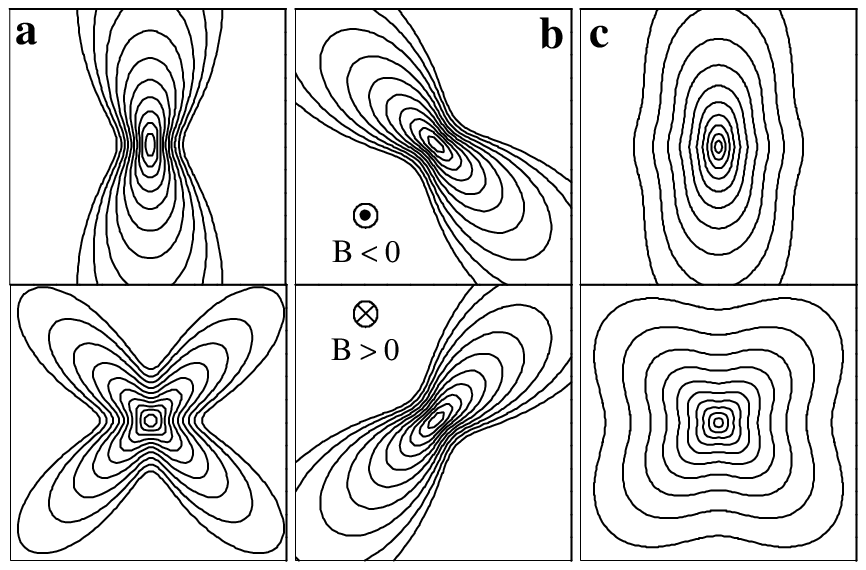}
\end{center}
\caption{Theoretical CBS cones (double scattering). (a) and (b) use the 
$J=0 \rightarrow J^{\prime}=1$ transverse plane model described in text
 (same parameters as in Fig.1). (c): cones in $lin\,{\parallel }\,lin$ (top)
 and $lin\,{\perp }\,lin$ (bottom) for a $J=3 \rightarrow J^{\prime}=4$
 transition and $B=0$ (semi-infinite medium); the cones shapes are rounded off
 compared to (a), but the symmetry properties are identical.}
\end{figure}
  
Figure 2 shows some examples of the double scattering CBS cones obtained in
the linear channels with this model. (a) and (b)
correspond to the same parameters as in Fig. 1. One clearly sees ((a), bottom
row) the distinctive clover-leaf shape in the $lin\,{\perp }\,lin$ channel
at $B=0$ with 4 lobes along the bisectors (at $45^{\circ }$). At 
non-vanishing magnetic field one of the lobe pairs is favored depending on
the sign of $\varphi $ (b). The comparison between Fig. 1 and 2 confirms that
the CBS cone tilt observed experimentally is consistent with the sign of the
dipole rotation, and that our simple model catches the essential part of the
physics. For off-resonant excitation $\delta _{\text{L}}\neq 0$, the picture
becomes more complex since the induced dipole is elliptical. However, if the
Zeeman splitting is large enough $\mu B\gg \Gamma $ and if the laser
frequency is tuned to resonance with one of the circular components 
$\delta _{\text{L}}=\pm \mu B$, then the dipole is essentially circular 
and the cone is isotropic in agreement with the data in Fig. 1(c).

We now discuss the effect of the internal structure in the ground state,
which is neglected in the previous model. Due to its internal structure, the rubidium atom
does not behave like a pure dipole scatterer. More specifically, 
the radiation pattern depends on the ground state magnetic sublevels 
$m_{g}$ and $m_{g}^{\prime}$ respectively before and after the scattering event.
To determine the total CBS signal, one has to sum over all possible transitions 
($m_{g1}\rightarrow m_{g1}^{\prime}$, $m_{g2}\rightarrow m_{g2}^{\prime}$) of two atoms. 
We have computed numerically the total
CBS signal in this case and found that, for moderate B values (up to a few Gauss), the 
symmetry of the CBS pattern is only weakly modified compared to the 
$J = 0 \rightarrow J^{\prime } = 1$ case. Indeed, in this regime, the CBS signal is 
dominated by the "Rayleigh" transitions $m_{g}\rightarrow m_{g}$, yielding the same 
behavior as in the $J = 0 \rightarrow J^{\prime } = 1$ situation. Thus, and this is the 
important point, we can still use the results of 
the simple $J = 0 \rightarrow J^{\prime } = 1$ model with an 
average dipole rotation $\varphi $ to describe the cone behavior for the 
$J = 3 \rightarrow J^{\prime } = 4$ transition of rubidium.
One illustration of the effect 
of the internal structure is given in Fig. 2(c) where we
plot the double scattering CBS signal at $B=0$ in the $lin\,{\parallel }\,lin$
(top) and $lin\,{\perp }\,lin$ (bottom) channels for the $J=3\rightarrow
J^{\prime }=4$ transition and a semi-infinite geometry. The main
effect of the internal structure in this case is a rounding-off of the cone features. 
The comparison with the experimental cones of Fig. 1(a) yields a nice agreement. 
In order to make a quantitative test of our model, we have experimentally
measured the dipole rotation $\varphi $ (open squares in Fig. 3(a)) by
analyzing the polarization of resonant light scattered by an optically-thin 
atomic cloud (optical thickness $0.05$). In
this regime, Faraday rotation and multiple scattering are negligible.
The solid line in Fig. 3(a) corresponds to the theory for a $J=3\rightarrow
J^{\prime }=4$ transition and agrees well with our measurements.
In a different experiment, we have measured the Faraday rotation 
{\it per mean free path} $\theta _{F}$ (solid squares in Fig. 3(a)) by recording, as a
function of $B$, the polarization rotation of light transmitted by a cloud
of fixed optical thickness $1$. When the magnetic field increases, the total
scattering cross-section decreases yielding an increasing mean-free path. 
Thus, in order to keep the optical thickness constant one has to
increase either the density or the size of the cloud. We emphasize that
the sign of the Faraday rotation is indeed opposite to that of the dipole
rotation (i.e. negative). We also note that the Faraday rotation per
mean-free path increases with the magnetic field due to the
increasing mean-free path \cite{meaculpa}. The dotted curve corresponds to
the theory for a 3 $\rightarrow $ 4 transition and is again in excellent
agreement. Finally, the triangles in Fig. 3(b) show the measured angle $\alpha _{f}$
for which the dominant lobe of the cone in the $lin\,{\perp }\,lin$ channel
flips by $90^{\circ }$, as a function of the magnetic field, compared with
our model taking into account the internal structure (line). The
measurement is unpractical above $B=4$ G due to poor cone contrast. The excellent
agreement between the experiment and our model on both the sign and magnitude of the 
magneto-optical rotation in the CBS signal indicates that the Faraday rotation through 
the "skin layer" is negligible in our configuration, and confirm the magnetic field-induced 
dipole rotation as the mechanism underlying the observed CBS anisotropies.

\begin{figure}
\begin{center}
\includegraphics[scale=1.0]{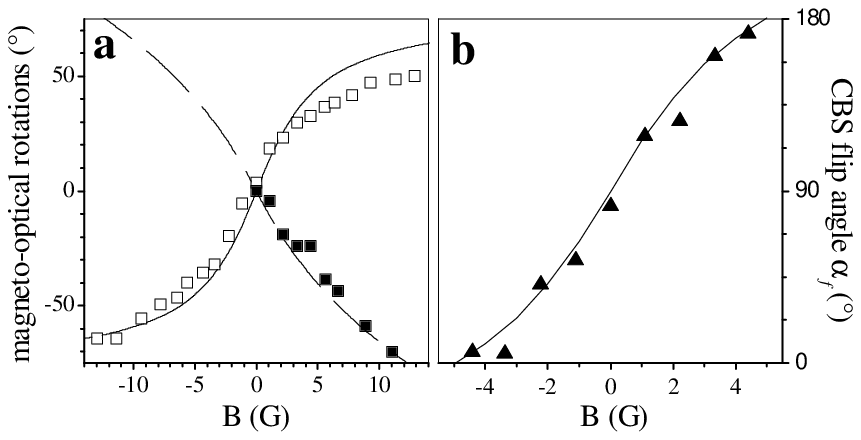}
\end{center}
\caption{Magneto-optical rotations. (a): measured dipole rotation 
$\varphi$ ($\square $) 
and Faraday rotation {\it per mean free path} $\theta _{F}$ ($\blacksquare $). 
The curves correspond to the theory for a $J=3 \rightarrow J^{\prime }=4$ 
transition (see text for sign 
conventions). Note that the two rotations have opposite signs. 
(b): measured CBS flip angle $\alpha _{f}$ ($\blacktriangle $) compared
to the $J=3 \rightarrow J^{\prime}=4$ model neglecting 
all propagation effects (solid line).}
\end{figure}

In conclusion, we reported in this paper the observation of various
anisotropies in the shape of the coherent backscattering cone from cold
rubidium atoms in a magnetic field. The observed behavior differs radically
from previous work on magneto-optically active samples where the cone
features were determined by the Faraday effect during the propagation
between scatterers. In our situation, the observed anisotropy is due to 
the modification of the atomic radiation pattern by the magnetic field. This
observation illustrates the new regimes that are accessible through the use
of cold atoms as a scattering medium for light. Some more elaborate modelling 
is needed to understand in detail the roles of propagation versus individual 
scattering effects in our sample, in particular to explain the behavior of the CBS 
enhancement factor in the presence of a magnetic field. 

We thank the CNRS and the\ PACA Region for financial
support. Laboratoire Kastler Brossel is laboratoire 
de l'Universit{\'e} Pierre et Marie
Curie et de l'Ecole Normale Sup{\'e}rieure, UMR 8552 du CNRS.


\begin{thebibliography}{9}
\bibitem{meaculpa}  contrarily to what was stated in \cite{Faraday}, where
the dependence of the scattering mean-free path with magnetic field was
overlooked.
\end{thebibliography}
\end{document}